\documentclass[letterpaper]{article} % DO NOT CHANGE THIS
\usepackage{aaai25}  % DO NOT CHANGE THIS
\usepackage{times}  % DO NOT CHANGE THIS
\usepackage{helvet}  % DO NOT CHANGE THIS
\usepackage{courier}  % DO NOT CHANGE THIS
\usepackage[hyphens]{url}  % DO NOT CHANGE THIS
\usepackage{graphicx} % DO NOT CHANGE THIS
\usepackage{natbib}  % DO NOT CHANGE THIS AND DO NOT ADD ANY OPTIONS TO IT
\usepackage{caption} % DO NOT CHANGE THIS AND DO NOT ADD ANY OPTIONS TO IT
\usepackage{algorithm}
\usepackage{algorithmic}

\usepackage{newfloat}
\usepackage{listings}
\DeclareCaptionStyle{ruled}{labelfont=normalfont,labelsep=colon,strut=off} % DO NOT CHANGE THIS
\floatstyle{ruled}
\newfloat{listing}{tb}{lst}{}
\floatname{listing}{Listing}
\title{INR-Based Generative Steganography by Point Cloud Representation}
\author{
    ZHONG Yangjie\textsuperscript{\rm 1,2},
    LIU Jia\textsuperscript{\rm 1,2}\equalcontrib,
    LUO Peng\textsuperscript{\rm 1,2},
    KE Yan\textsuperscript{\rm 1,2},
    CAI Shen\textsuperscript{\rm 3}   
    }

\affiliations{
     \textsuperscript{\rm 1}College of Cryptography Engineering, Engineering University of PAP, Shanxi, China\\
    \textsuperscript{\rm 2}Key Laboratory of Network and Information Security of PAP (Engineering University of PAP), Shanxi, China\\
    \textsuperscript{\rm 3}Donghua University, Shanghai, China\\
    1622177120@qq.com, liujia1022@gmail.com, lp\_nwpu@mail.nwpu.edu.cn, 15114873390@163.com, hammer\_cai@163.com
}

\begin{document}

\maketitle

\begin{abstract}
 Generative steganography (GS) directly generates stego-media through secret message-driven generation. It makes the hiding capacity of GS higher than that of traditional steganography, as well as more resistant to classical steganalysis. However, the generators and extractors of existing GS methods can only target specific formats and types of data and lack of universality. Besides, the model size is usually related to the underlying grid resolution, and the transmission behavior of the extractor is susceptible to suspicion of steganalysis. Implicit neural representation(INR) is a technique for representing data in a continuous manner. Inspired by this, we propose an INR-based generative steganography by point cloud representation (INR-GSPC). By using the function generator, the problem of the generator model size growing exponentially with the increase of gridded data has been solved. That is able to generate a wide range of data types and break through the limitation of resolution. In order to unify the data formats of the generator and message extractor, the data is converted to point cloud representation. We designed and fixed a point cloud message extractor. By iterating over the point cloud with adding small perturbations to generate stego-media. This method can avoid the training and transmission process of the message extractor. To the best of our knowledge, this is the first method to apply point cloud to generative steganography. Experiments demonstrate that the stego-images generated by the scheme have an average PSNR value of more than 65, and the accuracy of message extraction reaches more than 99\%.
\end{abstract}


\section{Introduction}
Image steganography is an information hiding method that uses images as stego-media to conceal secret messages. It aims to transmit secret messages covertly over public channels. At present, most image steganography methods adopt deep learning technology, with encode-decode structure as the mainstream method. A neural network is used as an encoder to embed the message into the cover-media to construct the stego-media. Another neural network is used as a decoder to extract the secret message. Baluja et al.\cite{Baluja2020HidingIW} and Zhu et al.\cite{Zhu2018HiDDeNHD} used this structure to construct steganography. With the development of deep generative models, the concept of Generative Steganography(GS) was proposed. GS is driven by the secret message to generate stego-media through a generator and uses a neural network as the message extractor. The traditional GS uses Generative Adversarial Networks(GAN)\cite{Goodfellow2021GenerativeAN,Skorokhodov2021StyleGANVAC,Anokhin2020ImageGW} or diffusion models\cite{Wei2023GenerativeSD,Ho2022ClassifierFreeDG,Ho2020DenoisingDP} as the generative model. 

However, both generators and extractors in GS still have obvious limitations. Firstly, cover-media or stego-media generated by the generator is represented by explicit data format(e.g, gridded images). This makes the parameter size of the generative model closely related to the resolution of the generated objects, showing an exponential growth relationship. This can lead to low efficiency in training the generator. Secondly, explicit data makes the current generative models only support generating single data type, such as images or videos. Moreover, the generated object has a fixed size (e.g, the resolution of images), making it difficult to construct a universal generation model. It limits the universality and flexibility of steganography. Furthermore, similar to the issues with generators, extractors can only deal with explicit data. Therefore, the extractor faces the problems of large parameter scale and single data type. More importantly, the current GS requires transmitting the message extractor to the receiver. This behavior can easily raise doubts about steganalysis, further increasing the transmission burden of secure channels and posing significant security risks.

As the Implicit Neural Representation(INR) technique has emerged, it has become an important trend of data representation. INR is a representation of a continuous function that takes gridded coordinates as input and returns features. Take images as examples, we can define a function $f:R^2\to\ R^3$ that maps pixel positions to RGB values using a neural network. This function $f$ is the INR of the image. This representation is independent of signal resolution.

Inspired by INR, some researchers try to introduce it into steganography. Liu et al.\cite{Liu2023HidingFW} realized steganography using function expansion after representing the data as INR. Dong et al.\cite{Dong2024ImplicitNR} completed steganography using model pruning method on the basis of INR. Similarly, Han\cite{Han2023DeepCS} hid secret data after converting it into INR. The use of INR extends the data types used for steganography and solves the problem of a single data type.

\begin{figure*}
  \centering
  \includegraphics[width=\textwidth]{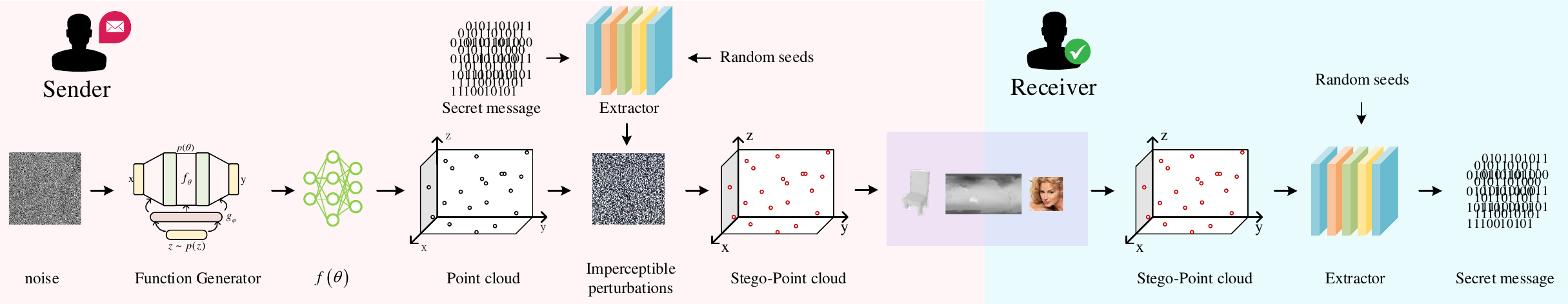}
  \caption{The whole framework, showing the scheme process from the sender and receiver.}
  \label{fig1}
\end{figure*}

Deep steganography also faces the similar extractor transmission problem as GS. To solve this problem, researchers have proposed two strategies. One is to disguise the message extractor. Li et al.\cite{Li2023SteganographyOS,Li2023TowardsDN} disguised the extractor as a neural network that performs other tasks to hide its real purpose. The other strategy is to avoid the transmission of the extractor, Zhang et al.\cite{Zhang2018GenerativeSB} used the Cardan Grille to determine the location of the message embedding and used the parameter as a shared key between the two parties. Kishore et al.\cite{Kishore2022FixedNN} and Luo et al.\cite{Luo2023SecuringFN} fixed the message extractor, and the two parties share the network structure of the extractor with the random seeds(also known as a key) used to initialize the network.

All of the above schemes solve the problems of data representation or extractor transmission individually, but have not been applied to GS. At present, there are still limitations in GS. In this paper, our goal is to solve the above problems. We consider constructing a universal and cover-independent generator in GS. And to design a message extractor that is also universal and can avoid the transmission process. In addition, efficiency and robustness should be considered in the designed scheme for practical deployment and real-world applications.

In order to achieve the above objectives, we construct GS based on INR and point cloud. The overall scheme is shown in Fig.\ref{fig1}. First, we use a function generator as the generative model, which extends the data types used for steganography and eliminates the strong coupling between the size of the generative model and the size of the data. Second, we sample point clouds from generated functions as the actual steganography object. Images, climate data, 3D shapes, etc. can be represented by point cloud, which unifies the data representation and improves the universality of the sheme. Then, we design and fix a point cloud message extractor for recovering secret messages. The receiver can recover the secret message by sharing the network structure of the extractor and random seeds. Thus, avoiding the transmission of message extractors.

The main contributions of this article include:

1. INR-Based Generative Steganography: For the first time, a generative function model by INR was used as the generator for GS, solving the problem of the generator model size increasing with data size. It can receive different types of multimedia data as stego-media, breaking through the resolution limitation of traditional gridded data.

2. Universal Data Representation: For the first time, point cloud representation is used as the actual steganographic object in GS. Distinguishing it from function data and gridded data, point cloud provides a unified data format for steganography. It extends the domain of steganography and improves the universality.

3. Point Cloud Message Extractor: Designing and fixing a point cloud message extractor. It solves the problem that the traditional extractor can only target fixed kinds of data, and extends the universality of the message extractor. At the same time, it avoids the transmission behavior.

\section{Related Work}

\subsection{Steganography By Deep Generative Models}
Steganography by deep generative models can be categorized into the following three types based on the application of generative models in steganography:

\subsubsection{Cover-media Generation}
Early in deep cryptography, some scholars used generative models to directly generate the cover-media. GAN is one of the most commonly used models. The earliest introduction of GAN into steganography mainly utilizes the adversarial strategy to generate realistic cover images, such as SGAN\cite{Volkhonskiy2017SteganographicGA} and SSGAN\cite{Shi2017SSGANSS}.

\subsubsection{Modification Strategy Generation}
In addition to directly generating the cover-media, researchers also use modification strategy generation in steganography. ASDL-GAN\cite{Tang2017AutomaticSD} and UT-SCA-GAN\cite{Yang2018SpatialIS} generated probability modification matrices using adversarial processes within the framework of minimizing distortion costs. 

\subsubsection{Stego-media Generation}
Another way is to use deep generative models to directly generate stego-meida. One method is to generate stego-media based on message mapping. Liuet al.\cite{YYKX201802015} replaced the class labels of GANs with secret messages as the driving factor to directly generate stego-media. Hu et al.\cite{Hu2018ANI} directly generate stego-media. Another method is to train the message extractor to directly extract messages from the generated cover by defining a loss function. Liu et al.\cite{Liu2018DigitalCG} proposed a generative steganography based on constraint sampling. Zhang et al.\cite{Zhang2021PixelStegaGI} used autoregressive models and arithmetic coding algorithms to achieve pixel-level steganography, adaptively embedding secret messages based on pixel entropy to generate stego-media.

However, the application of the above three generative models in steganography still faces the following issues: firstly, the data generated by the generator is still explicit representation, resulting in strong coupling between model size and data size. Taking images as an example, the parameter size of the generator increases exponentially with resolution. The second is that the generator generates objects with a single type and fixed size. A trained generator can only support generating fixed data, such as image data of a certain resolution.

\subsection{Steganography By Fixed Extractor}
In GS, the message extractor needs to be transmitted to the message receiver. But there is a risk of exposing steganographic behavior. There are two strategies that can be adopted to reduce its security risks. One strategy is to directly avoid the transmission behavior of the message extractor. Liu et al.\cite{Liu2018DigitalCG} tried to use the Cardan Grille for message embedding location to avoid the transmission of the message extractor. Kishore et al.\cite{Kishore2022FixedNN} proposed Fixed Neural Networks Steganography(FNNS) to avoid the transmission of the message extractor. Luo et al.\cite{Luo2023SecuringFN} added the key based on \cite{Kishore2022FixedNN} to improve the security.

Another strategy is to disguise the message extractor. Li et al.\cite{Li2023SteganographyOS,Li2023TowardsDN} disguised the steganographic network as an ordinary deep neural network. Although these two approaches ensure the security of the message extractor, they still cannot avoid the problem of high communication burden for the transmission of the extractor.

To our knowledge, there is currently no solution to use fixed message extractors in generative steganography. We design a fixed message extractor specifically for point cloud and apply it to generative steganography, expanding the universality of the message extractor and avoiding the risk of steganalysis suspicion caused by extractor transmission. However, neither the fixed message extractor system nor the use of different generative models has solved the strong correlation between model size and data size. The data types are also limited. The universality and flexibility still need to be improved.

\subsection{INR-Based Steganography}

\subsubsection{Implicit Neural Representation}
INR refers to defining images or any neural network as a function\cite{Ha2016GeneratingLI}. It is a new method for parameterizing various signals (images, audio, video, 3D shapes). It parameterizes the signal into a continuous function, mapping the domain of the signal to the values of attributes at that coordinate, also known as coordinate-based representation. This representation method no longer relies on traditional data formats such as grids or voxels. It uses neural networks themselves as the representation of data, transforming explicit data into implicit data. Recent work has demonstrated the potential for continuous and memory-efficient implicit representations of shapes\cite{Genova2019LearningST}, objects\cite{Park2019DeepSDFLC,Michalkiewicz2019ImplicitSR}, or scenes\cite{Sitzmann2019SceneRN,Jiang2020LocalIG} as functions.

\subsubsection{INR-Based Steganography}
The emergence of INR provides a new representation of data for steganography. Han et al.\cite{Han2023DeepCS} implicitly represented multimedia data and then sampled the cover-image for steganography. But its essence is still multimedia steganography. Liu et al.\cite{Liu2023HidingFW} firstly proposed the concept of INR-Based steganography. They performed the secret message as INR before expanding the function. Yang et al.\cite{Song2023ImplicitSB} disarranged multimedia after representing it as INR and mixed it into a large network for steganography. But the disarranged network only outputs noisy data, which is low in covertness for steganography. Dong et al.\cite{Dong2024ImplicitNR} realized steganography by INR by finding the redundant space and using model pruning, magnitude-based weight selection, and secret weight replacement method. However, the above schemes essentially use cover modification strategies to construct stego-media, just transferring to the domain of the function. Besides, INR has not yet been applied in GS.

\subsubsection{INR-Based Generative Model}
In order to apply INR to GS, we need a generator that can directly generate functions. Currently, there are the following studies on function generators: Dupont et al.\cite{Dupont2021GenerativeMA} abandoned discrete grids and used continuous functions to parameterize individual data points, and then constructed a generative model by learning distributions of the functions. Parket et al.\cite{Park2024DDMIDL} proposed a diffusion model for INR domains. This model generates adaptive positions rather than embedding by the neural network's weights to achieve high-quality generation. Dupont et al.\cite{Dupont2021GenerativeMA} proposed to train a base model followed by data to INR to improve the efficiency of generative models\cite{Wei2022GenerativeSN}. 

Inspired by INR, we first use a function generator in GS. It replaces traditional explicit representation with implicit representation. It eliminates the strong coupling between model size and resolution size in the traditional scheme, and extends the types of cover-media used for steganography. In order to make the extractor match with the function generator, we design and fix a point cloud message extractor inspired by FNNS. The advantage of this approach is that it makes the model size of the extractor independent of the data size and type. By innovating the generator and the message extractor, we make our scheme applicable to different types of multimedia data, extend the cover-media types of steganography, and improve the universality and flexibility of the scheme.

\begin{figure}[h]
  \centering
  \includegraphics[width=\columnwidth]{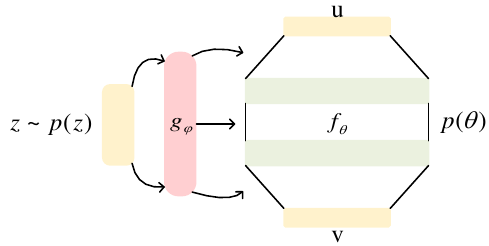}
  \caption{Function Generator.}
  \label{fig4}
\end{figure}

\begin{figure*}
  \centering
  \includegraphics[width=\textwidth]{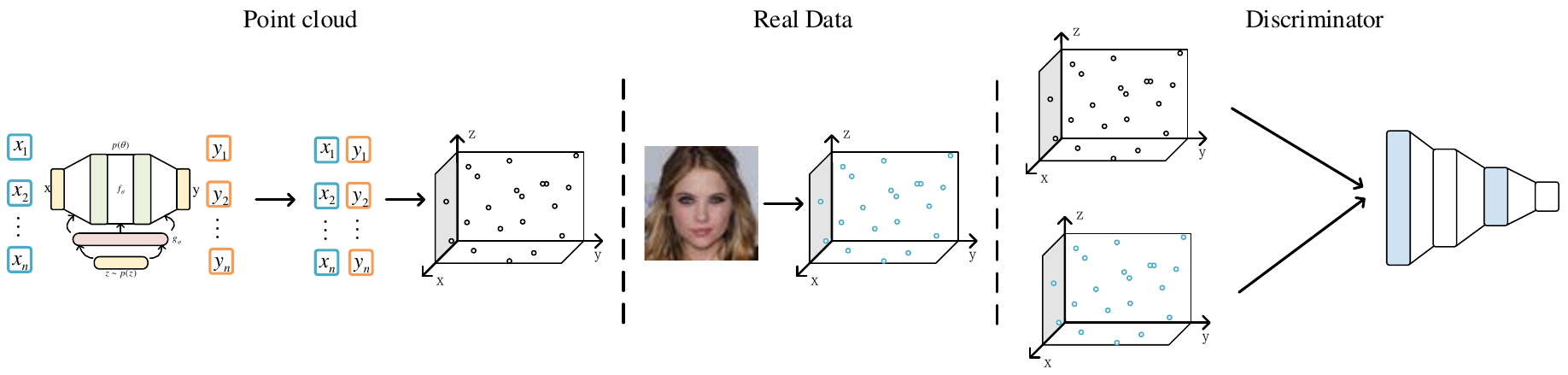}
  \caption{The Training Process of Function Generator.}
  \label{training}
\end{figure*}

\section{The Proposed Method}
The framework is shown in Fig.\ref{fig1}. Unlike the explicit representation of data in traditional steganography, we use the implicit representation of data as the processing object. The method can be divided into two parts: cover-media generation and stego-media construction. We generate the cover-media(function) through a function generator. Since point cloud data is the most commonly used representation of INR, we sample the point cloud from the function as the actual steganographic object. Finally we design and fix a message extractor for the point cloud to construct stego-media. The detailed steps are described below.

\subsection{Cover-media Generation}

\subsubsection{Data Representation}
Traditional multimedia data is represented through gridded data, which converts continuous signals into discrete signals and represents them in the form of pixels. The disadvantage of this method is the loss of detailed information about the data, and the limited types of data that can be represented. Point cloud is independent of data types, constructing a unified data format. Wu et al.\cite{Wu2018PointConvDC} believed that it is inappropriate to view the coordinates of an image as scattered and unrelated discrete points. This view ignores the concept of potential distance between coordinates. 

Taking images \textit{I} as an example, the position of each pixel corresponds to a coordinate value $u=(x,y)$ and an RGB value $v=I(x,y)$. The coordinates and features are considered as a set $((x,y), I(x,y))$, and such a set is called a point cloud pair. Assuming $\{(u_{i},v_{i})\}_{i=1}^{n}$ is a set of point clouds corresponding to all pixel positions and RGB values. The implicit representation of this image can be represented by minimizing the function:
$$min_\theta\sum_{i=1}^n\|f_\theta(u_i)-v_i\|_2^2$$

Among them, $v_i=f_{g_\varphi(z)}(\gamma(u_i))$,$\gamma$ represents random Fourier feature (RFF) encoding\cite{Tancik2020FourierFL}, used to learn high-frequency information of the function. Represent coordinates with $\mathrm{u}\in\mathrm{U}$, features with $\mathrm{v}\in\mathrm{V}$, and all data that can be represented by $(u,v)$ can be represented by INR. This representation of INR based on point cloud abandons traditional gridded data and directly operates on coordinate and feature pairs. Actually, the point cloud is a constant representation of INR. The point cloud can be converted to a continuous representation of INR, and INR can also be obtained from the point cloud by dense sampling. Therefore, a universal data representation is constructed by choosing to use the point cloud as the stego-media. In turn, the point cloud can be represented as explicit data by data converter.

\subsubsection{Function Generator}
The purpose of this paper is not only to represent the data as INR. We want to apply the INR to generative steganography. So we need a function generator for continuously generate function instances. To generate continuous functions, researchers have provided a number of methods\cite{Du2021LearningSM,Zhuang2023DiffusionPF,Koyuncu2023VariationalMO}. Among others, Dupont et al.\cite{Dupont2021GenerativeMA} proposed a method to construct function generators by learning function distributions for function modeling. In this paper, we use function generator as the generative model.

proposed Generative Adversarial Stochastic Process(GASP) to construct a function generator and applied point cloud processing to the process. This is similar to our idea. Based on this, we use this method to construct generators for steganography.

The network architecture of the function generator is illustrated in Fig.\ref{fig4}. Assuming that the network structure of a single data point represented by function $f_{\theta}$ is fixed, learning the distribution of function $f_{\theta}$ is actually learning its weight $p(\theta)$. It is determined by the latent vector $p(z)$ and the function generator $g_{\varphi}{:}Z\to\Theta$. $g_{\varphi}$ maps the latent vector space to the weights in function $f_{\theta}$ through parameter, and a series of weights $\theta=g_{\varphi}(z)$ can be obtained by sampling from $p(z)$ through \textbf{G}. The purpose of the generator is to learn the function $f_{\theta}$ by training, and the purpose of learning the function is to get the distribution of the function. But usually there is no access to the specific function expression, and all that can be obtained are data points that match the distribution of the function.

In order to train the function generator, we refer to Generative Adversarial Stochastic Process(GASP) by Dupont et al.\cite{Dupont2021GenerativeMA} to learn the distribution of the function. GASP define a discriminator to ensure the authenticity of the generated functions. The training process of the discriminator is shown in Fig.\ref{training}. The point clouds sampled from the generated function are input into the discriminator together with the point clouds generated from the real image after data conversion. The generated point cloud pairs are judged to be real or not by GASP. Finally, the function is output.

During the training process, GASP defined a penalty $R_1$ equivalent to the regularization proposed by Mescheder\cite{Mescheder2018WhichTM}, which is oriented towards point cloud. 
And we use traditional GAN loss for training. For images, the regularization $R_1$ corresponds to penalizing the gradient norm of the discriminator with respect to the input image. After representing the image as point cloud representation, for a set $\{(u_i,v_i)\}_{i=1}^n$, we define the penalty as follows:

\begin{equation}
R_{l}(\mathbf{s})=\frac{l}{2}\parallel\nabla_{\mathbf{v}_{l},\cdots,\mathbf{v}_{n}}D(\mathbf{s})\parallel^{2}=\frac{l}{2}\sum_{\mathbf{v}_{i}}\parallel\nabla_{\mathbf{v}_{i}}D(\mathbf{s})\parallel^{2}
\label{eqa1}
\end{equation}

Here, \textbf{D} represents the loss of the discriminator. In this process, we penalize the gradient of the discriminator using feature values. The whole process gets rid of the discretized gridded data and works directly on point cloud. We generate cover-media through this process.

\subsection{Stego-media Construction}

\begin{figure}[h]
  \centering
  \includegraphics[width=\columnwidth]{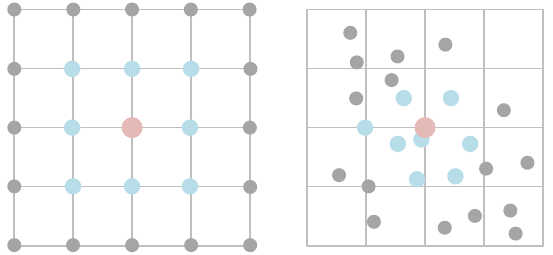}
  \caption{Convolution Neighborhood for Regular Convolutions and PointConv.}
  \label{fig3}
\end{figure}
\subsubsection{Point Coud Message Extractor}

Traditional steganography is always based on gridded data. So they use convolutional networks. But in this paper, we use point cloud as cover-media, which are different from the neat and regular arrangement of traditional gridded data. Actually, the arrangement of point cloud is disordered and in the concept of potential distance, as shown in Fig.\ref{fig3}. So the traditional message extractor for grids is no longer applicable. Wu et al.\cite{Wu2018PointConvDC} proposed the PointConv computational framework based on the specificity of point cloud arrangement. Based on this, we design message extractors specialized for point cloud and use point cloud convolution in the process. It can enable the message extractor to match the data type of the generator. It is able to adapt to different types of data and improve the universality of the extractor.

\subsubsection{Fixed Point Cloud Message Extractor}
Deep neural networks have a highly sensitive characteristic to small changes in input. Based on this feature, Kishore et al.\cite{Kishore2022FixedNN} used the fixed message extractor. They used an image with imperceptibly small perturbations, constantly iterating over the image to achieve secret message embedding.

\begin{figure}[h]
  \centering
  \includegraphics[width=\columnwidth]{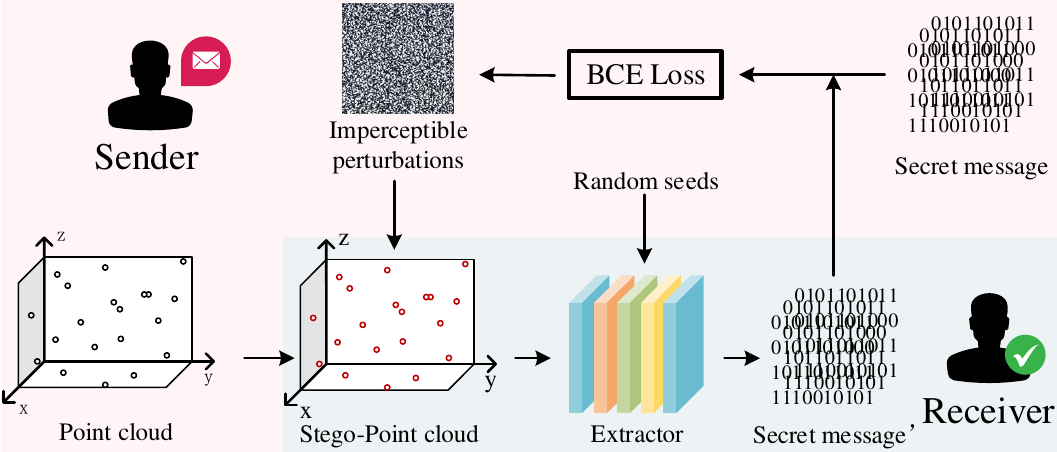}
  \caption{Training Process of Fixed Point Cloud Message Extractor.}
  \label{fig6}
\end{figure}

Inspired by Kishore\cite{Kishore2022FixedNN}, we fix the point cloud message extractor. Compared with traditional steganography, the advantage of the fixed point cloud message extractor is that there is no longer a need to pass the pre-trained message extractor. Both the sender and receiver only need to share the neural network with the same network structure and the random seed used to initialize the network to achieve the extraction of the secret message. The object of training is changed from the extractor to the point cloud itself. The process is shown in Fig.\ref{fig6}. The sampled point cloud is fed into the fixed message extractor, and small perturbations are added to the point cloud to realize the construction of the stego-point cloud. Calculate the BCE loss of the extracted secret message from the real secret message and modify the perturbation until the loss is minimized. The whole process avoids the delivery of the message extractor, which improves the security and reduces the training cost at the same time.

\subsubsection{Loss Function}
During the training process for generating stego-point clouds, we sample point clouds from the function and input them into a fixed message extractor. After comparing the extracted secret message with the original secret message, we add slight perturbations to the point cloud. We use an unconstrained L-BFGS algorithm\cite{Fletcher1988PracticalMO} to optimize the point cloud, with the defined message extraction loss as follows:
$$loss=-\frac{1}{N}\sum_{i=1}^{N}(m_{real}[i]log(m_{ext}[i])+(1-m_{real}[i])log(1-m_{ext}[i]))$$

Where N is the batch size, $m_{real}$is the real secret message, and $m_{ext}$ is the extracted secret message.

\section{Experiments}
To facilitate a clear visual analysis of the experimental results, we visualize the generated point cloud as two-dimensional images using a data converter.

\subsection{Evaluation Metrics}
We evaluate its steganographic capability from various aspects, including visual security, image quality, message extraction accuracy, undetectability, indistinguishability, robustness, efficiency, and ablation experiments. Additionally, we assess the special capability of super-resolution sampling in light of its data representation continuity. We define the following criteria:

Extraction accuracy(Acc):
$$acc=1-BER(m,m^{\prime})$$
$$=1-\frac{\sum_{l=1}^{L_m}XOR[m(l),m'(l)]}{L_m}$$
Where BER is the Bit Error Rate; XOR is the exclusive OR operation, used to calculate the number of erroneous bits; $m(l)$ is the original secret message; $m^{\prime}(l)$ is the extracted secret message; and $L_{M}$ is the length of the secret message.

Unstructured pruning rate($prune\_ratio$):
$${\mathrm{prune\_ratio}}={\frac{num\_to\_prune}{num\_element}}$$
Where $num\_to\_prune$ represents the number of elements to be pruned, and $num\_elements$ represents the total number of elements.

Peak Signal-to-Noise Ratio(PSNR):
$$\mathbf{MSE}=\frac{1}{HW}\sum_{i=1}^{H}\sum_{i=1}^{W}\left[\mathbf{X}_{i,j}-\tilde{X}_{i,j}\right]^{2}$$
$$\mathbf{PSNR}=20\mathrm{log}_{l0}(max_{\mathrm{X}})-10\mathrm{log}_{l0}(\mathbf{MSE})$$

Peak Signal-to-Noise Ratio is used to measure the difference between the original image \textbf{X} and the stego-image $\tilde{X}$. It is a commonly used metric for assessing image quality. Let H and W represent the resolution of the images. We use PSNR to evaluate the quality of images before and after steganography.

Structural Similarity Index(SSIM):
$$\mathbf{SSIM}=\frac{(2\mu_{\mathrm{X}}\mu_{\mathrm{\tilde{X}}}+c_{1})\left(2\sigma_{\mathrm{X}}\tilde{\mathrm{X}}+c_{2}\right)\left(\sigma_{\mathrm{X}}^{2}+\sigma_{\mathrm{\tilde{X}}}^{2}+c_{2}\right)}{\left(\mu_{\mathrm{X}}^{2}+\mu_{\mathrm{X}}^{2}+c_{1}\right)\left(\sigma_{\mathrm{X}}^{2}+\sigma_{\mathrm{\tilde{X}}}^{2}+c_{2}\right)}$$
Where $c_{1},c_{2}$ are very small stability constants. SSIM is used to measure the similarity between the original image \textbf{X} and the stego-image$\tilde{X}$. The difference between PSNR and SSIM is that PSNR considers pixel-level differences, while SSIM takes structural differences into account.

\subsection{Setup}
Our experiments were conducted on an NVIDIA GeForce RTX 4060 Ti graphics card. The experimental environment utilized PyTorch version 2.5.1, torchvision version 0.20.1, and CUDA version 12.1. For the images, we chose the Celebahq dataset for the experiments, and to ensure effective training, the images in the dataset were uniformly resized to a specific dimension. The climate data used for the experiments was sourced from the ERA5 dataset.

\subsection{Visual Security}
\begin{figure}[h]
  \centering
  \includegraphics[width=\columnwidth]{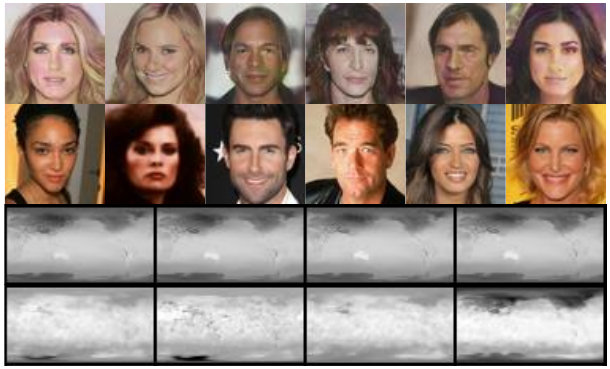}
  \caption{Visual effect demonstration.}
  \label{fig7}
\end{figure}

We conducted experiments on both the Celebahq and ERA5 datasets. For the image experiments, we randomly selected 6 results from 1000 trials for display, as shown in the first two rows of Fig.\ref{fig7}. The first row contains the generated stego-media, while the second row contains the real images. For the climate data, we selected four instances for display, demonstrating that the visually generated stego-media from the model are nearly indistinguishable from the real ones. This demonstrates the feasibility of our scheme in the diversity data representation.

\subsection{Image Quality}
This paper uses PSNR and SSIM as two metrics to evaluate the image quality.

\begin{figure}[h]
  \centering
  \includegraphics[width=\columnwidth]{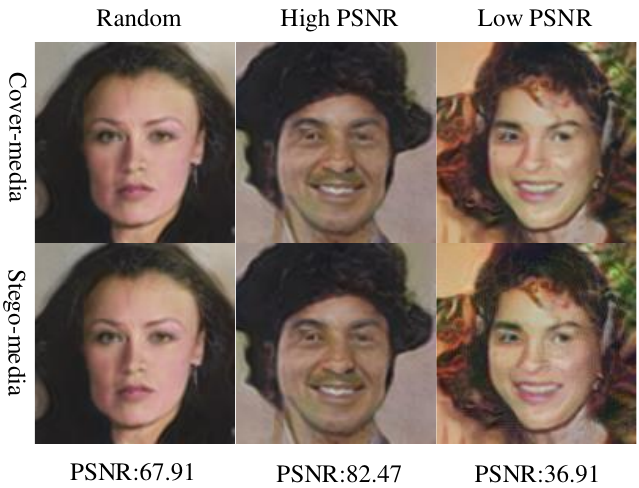}
  \caption{Samples with Random, High, and Low PSNR Values.}
  \label{fig8}
\end{figure}

Fig.\ref{fig8} presents the experimental results with a resolution of 128×128. The leftmost column shows the PSNR values generated by random sampling from the model. The middle and rightmost columns showcase a set of experiments filtered based on higher and lower PSNR values, respectively. The images before and after steganography are difficult to distinguish with the naked eye. The PSNR values further demonstrate that INR-GSPC achieves extremely high image quality. Table.\ref{tab2} compares the PSNR and SSIM values of INR-GSPC with other steganography, indicating that INR-GSPC significantly outperforms other methods in terms of image quality.

\begin{table}
  \caption{Image Quality of Different Methods.}
  \label{tab2}
  \begin{tabular}{c|c|c}
    Method&PSNR&SSIM\\
    SteganoGAN\cite{Zhang2019SteganoGANHC} & 25.98& 0.85\\
    FNNS-D\cite{Kishore2022FixedNN} & 36.06& 0.87\\
    FNNS-R\cite{Kishore2022FixedNN} & 39.79 & 0.76\\
    Key-based FNNS\cite{Luo2023SecuringFN} & 39.48& 0.95\\
    \textbf{INR-GSPC} &66.99&0.99\\
\end{tabular}
\end{table}

\subsection{Accuracy}
If time is not restricted during the experimental process, a 0\% message extraction error rate can be achieved through continuous modifications of the point cloud. In actual experiments, to improve the efficiency of the scheme, a condition is set to end the loop when the message extraction error rate is below 1\%. Table.\ref{tab3} demonstrates the message extraction error rates of different schemes when the embedded message capacity is 1 bpp.

\begin{table}
  \caption{Error Rate of Different Methods}
  \label{tab3}
  \begin{tabular}{c|c}
    Method&Error Rate\\
    SteganoGAN\cite{Zhang2019SteganoGANHC} & 3.94\% \\
    FNNS-D\cite{Kishore2022FixedNN} & 0\%\\
    FNNS-R\cite{Kishore2022FixedNN} & 0.14\%\\
    Key-based FNNS\cite{Luo2023SecuringFN} & 3E-04\%\\
    StegaDDPM\cite{Peng2023StegaDDPMGI} & 7.55\%\\
    LDStega\cite{Peng2024LDStegaPA} & 1.35\%\\
    \textbf{INR-GSPC} &<1\%\\
\end{tabular}
\end{table}

\subsection{Undetectability}
\begin{figure}[h]
  \centering
  \includegraphics[width=\columnwidth]{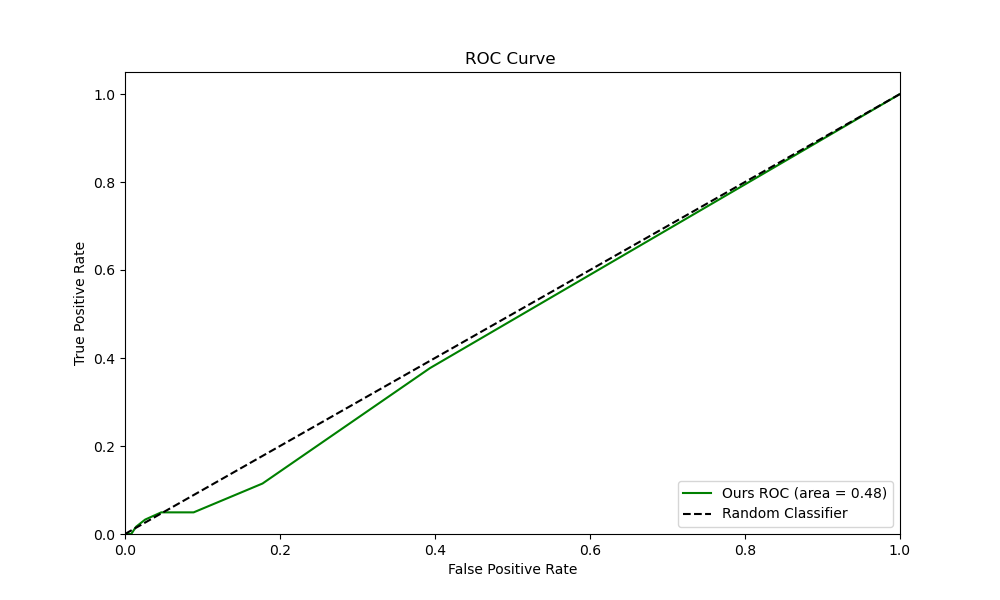}
  \caption{ROC.}
  \label{fig9}
\end{figure}

We use the open-source steganalysis tool StegExpose\cite{Boehm2014StegExposeA} for the analysis. A test set is established with a folder containing 1000 stego-images with a resolution of 64×64. StegExpose detected 76 images as stego-images, resulting in a detection rate of 7.6\%, as shown in Table.\ref{tab4}. Here, "Sample Pairs" refers to the set of image features used to compute the hidden information. "RS analysis" indicates randomness analysis.  "Fusion mean" represents the process of combining results from different methods to derive an average value. A lower Fusion value indicates a lower probability of hidden information existing within the image file, which also confirms the security of the scheme. When 1000 stego-images were mixed into a dataset of 140,000 real images, the area under the ROC curve was calculated to be 0.48, as shown in Fig.\ref{fig9}. This demonstrates that our scheme possesses a certain degree of security, allowing it to evade standard steganalysis tools.

\begin{table}
  \caption{Image Quality of Different Methods.}
  \label{tab4}
  \begin{tabular}{c|c|c|c}
    Detection rate&Sample Pairs&RS analysis&Fusion (mean)\\
    7.6\% &0.07776 &0.0808 &0.0892\\
\end{tabular}
\end{table}

\subsection{Indistinguishable}
To assess the indistinguishability between the stego-images $\mathrm{I}_{\mathrm{stego}}$ and the cover images $\mathrm{I}_{\mathrm{cover}}$, we employed the method of residual images for comparison, calculated through the following formula for the difference between the two images:

\begin{equation}
    R=|\mathbf{I_{cover}}-\mathbf{I_{stego}}|
\end{equation}

\begin{figure}[h]
  \centering
  \includegraphics[width=\columnwidth]{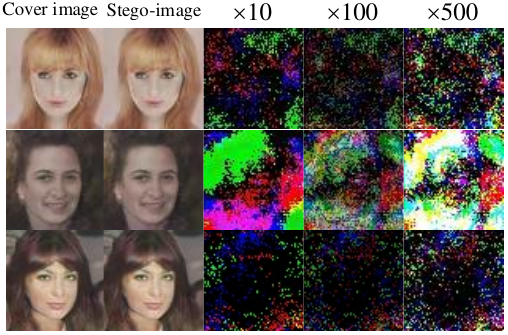}
  \caption{Comparison of Residual Images.}
  \label{fig10}
\end{figure}

The difference values were magnified by factors of 10, 100, and 500, as shown in Fig.\ref{fig10}. A smaller residual value indicates that the cover-image is visually more similar to the stego-image, demonstrating that our scheme performs well in terms of indistinguishability.

\subsection{Robustness}
\begin{figure}[h]
  \centering
  \includegraphics[width=\columnwidth]{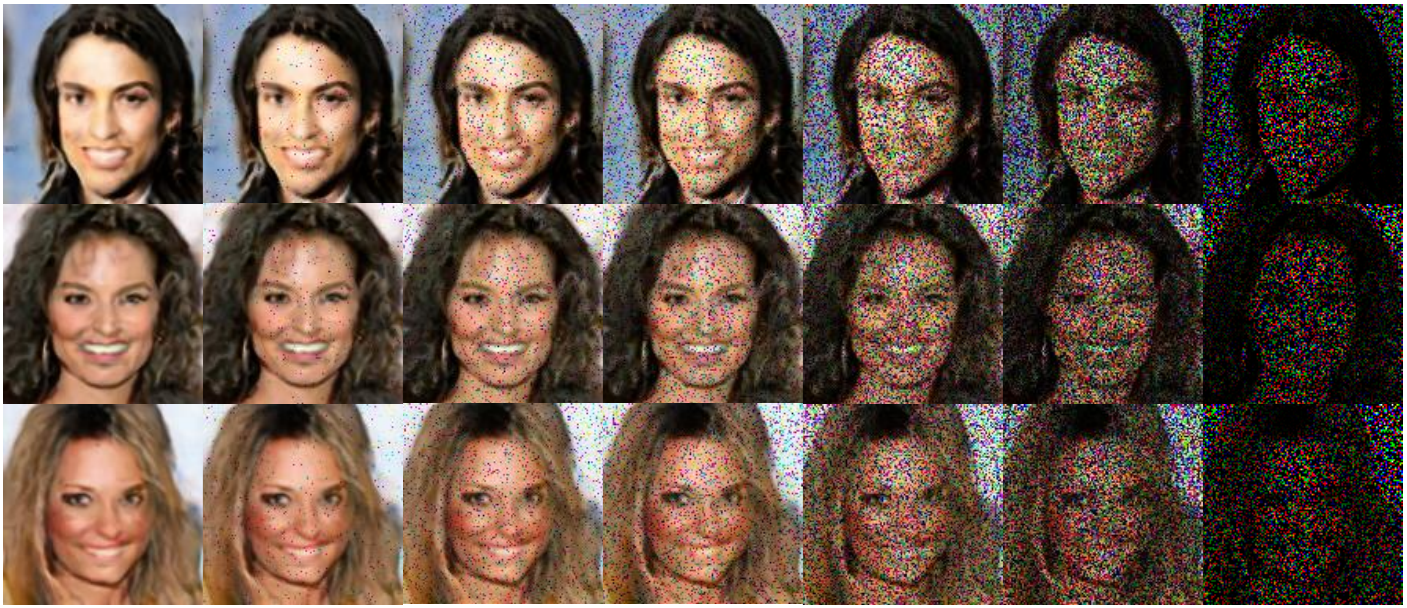}
  \caption{Images with Different Pruning Rates.}
  \label{fig11}
\end{figure}

Stego-media are inevitably compressed or noisy during transmission due to the channel, so it is necessary to consider their robustness. Since a continuous function is used during the training process, and the function acts as a type of neural network structure, traditional methods of evaluating robustness through noise addition are no longer applicable. For the point cloud, we used unstructured pruning to evaluate its robustness. Fig.\ref{fig11} provides a visual representation at a resolution of 128 for different deactivation rates. It is evident that at a deactivation rate of 0.8, the contours of the stego-media can still be recognized.

\subsection{Ablation Study}
\begin{figure}[h]
  \centering
  \includegraphics[width=\columnwidth]{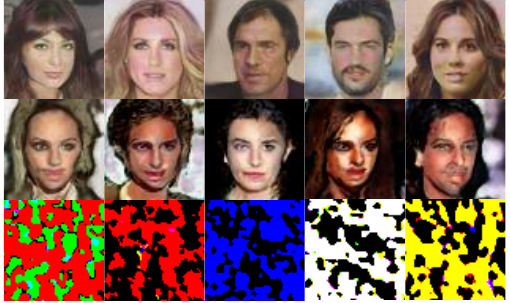}
  \caption{Ablation Experiment.}
  \label{fig12}
\end{figure}

In the aforementioned experiments, we set the learning rate of the function generator to 1e-4. In this section, we present the results for different learning rates, as shown in Fig.\ref{fig12}. From the results, it is evident that when the number of training epochs is the same, selecting a learning rate of 1e-4 as a hyperparameter yields the best visual effects.

\begin{figure}[h]
  \centering
  \includegraphics[width=\columnwidth]{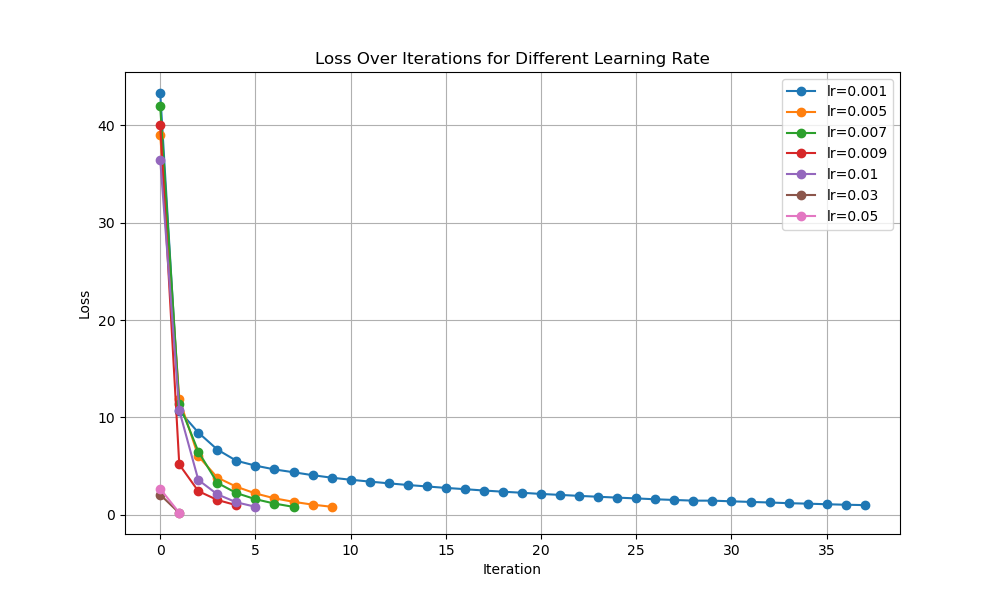}
  \label{fig13}
\end{figure}

To test the relationship between the extractor's learning rate and steganographic efficiency, we conducted tests on the number of iterations of the loss function at learning rates of 0.001, 0.005, 0.007, 0.009, 0.01, 0.03, and 0.05. The results are shown in Fig.\ref{fig13}. It can be observed that as the learning rate decreases, the iterations of the loss function become slower, requiring more epochs. Conversely, as the learning rate increases, the loss function decreases more rapidly, allowing for quicker achievement of the preset standards, thereby completing the steganographic process.

\subsection{Efficiency}
\begin{figure}[h]
  \centering
  \includegraphics[width=\columnwidth]{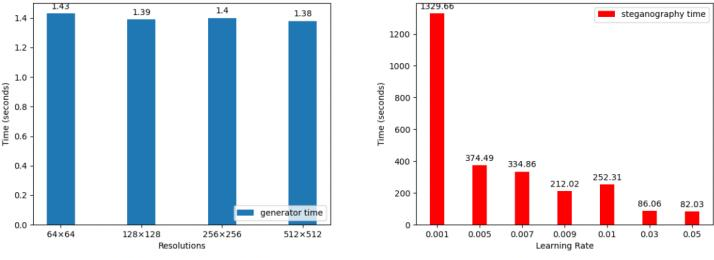}
  \caption{Efficiency Comparison.}
  \label{fig14}
\end{figure}

When the message embedding rate is 1 bpp, the training time for the generative model over 10,000 epochs was 24 hours. After training is completed, both the message hiding and embedding processes can be executed within a few seconds. The blue bars in Fig.\ref{fig14} represent the average generation time of stego-images at different resolutions. It can be observed that as the resolution increases, the time for generating stego-images does not increase exponentially as in traditional gridded representations, but instead remains relatively stable. The red bars in Fig.\ref{fig14} indicate that when the learning rate is below 0.1, the efficiency of the steganographic process is lower. However, as the learning rate increases, efficiency gradually improves. Combining this with the previous figure, selecting a learning rate of 0.03 as a hyperparameter for the extractor can better balance image quality and steganographic efficiency.

\subsection{Super-resolution}
\begin{figure}[h]
  \centering
  \includegraphics[width=\columnwidth]{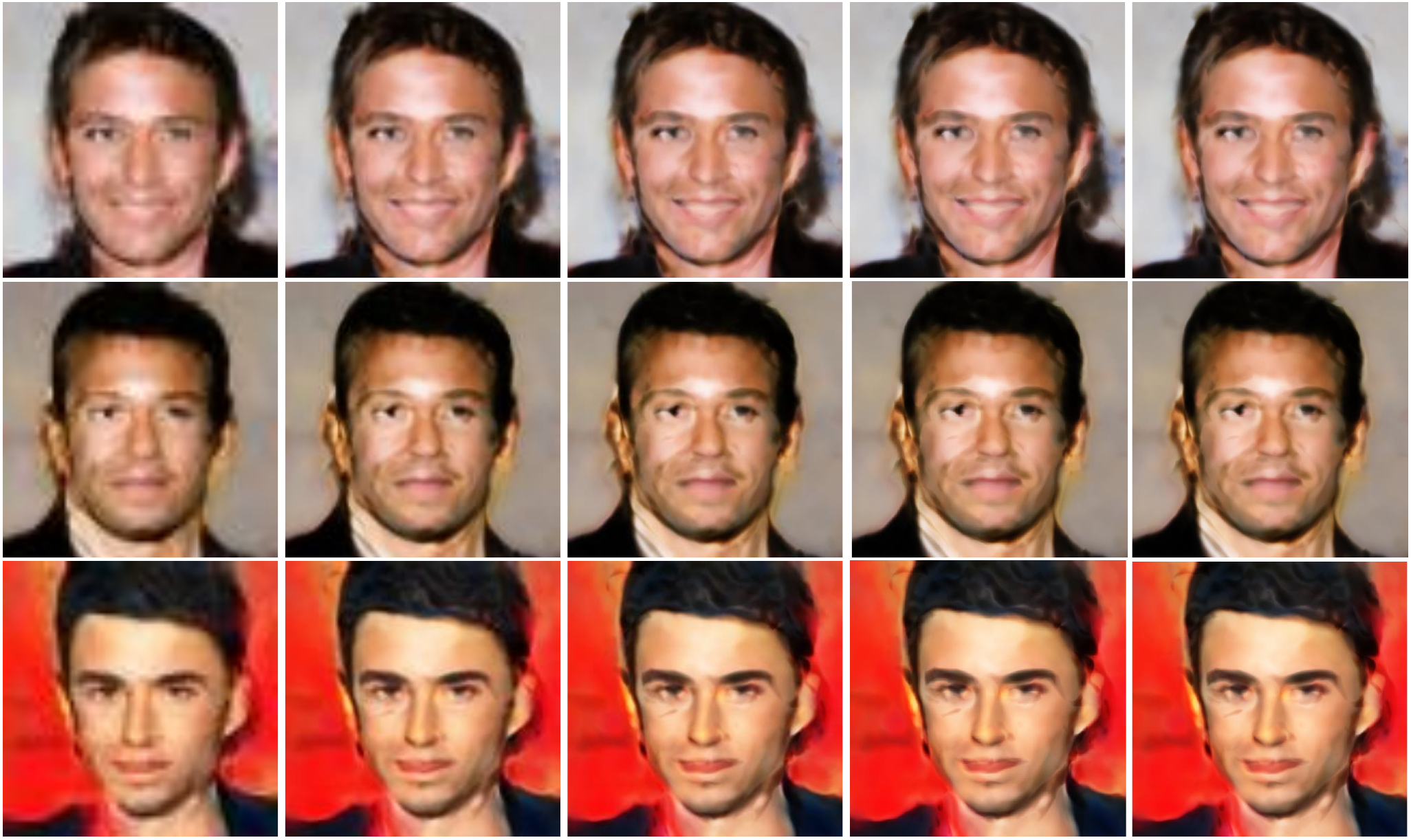}
  \caption{Comparison of Different Resolution.}
  \label{fig15}
\end{figure}

Fig.\ref{fig15} illustrates the stego-images generated through super-resolution sampling based on a model trained at a resolution of 64x64. The image sizes increase from 64, 128, 256, 512, to 1024 resolutions. In our approach, we utilize implicit neural representation methods to encode the images. This method has the property of being resolution-independent, allowing the generated images to not be constrained by a specific number of pixels or resolutions. But rather to be achieved through encoding and decoding via vectors in a high-dimensional implicit space. Therefore, we can sample stego-images from models trained at a fixed resolution at any resolution, providing high flexibility and generalization ability for processing image data at different resolutions.

\section{Conclusion}
This paper introduces point cloud representation as a steganographic object for the first time, constructing an INR-based Generative Steganography by Point Cloud Representation (INR-GSPC). This approach allows the model size to no longer grow exponentially with data resolution, while also applying to various types of multimedia data. Taking images as an example, it enables multi-resolution sampling, breaking through the limitations of resolution. We have designed and fixed a message extractor specifically for point cloud. Fixed extractor transfers the training of the extractor to point cloud, reduces training costs and mitigates the risks associated with extractor transmission. Experiments have demonstrated that the quality of the stego-images generated by our scheme is significantly higher than that of existing image steganography, with an average PSNR value of over 65 for the generated stego-images and an accuracy rate of over 99\% for message extraction.

\bibliography{sample-base}

\end{document}